\documentclass[letterpaper]{article} 
\usepackage{aaai25}  
\usepackage{times}  
\usepackage{helvet}  
\usepackage{courier}  
\usepackage[hyphens]{url}  
\usepackage{graphicx} 
\usepackage{natbib}  
\usepackage{caption} 
\usepackage{algorithm}
\usepackage{algorithmic}
\usepackage{array}

\usepackage{xcolor}
\newcommand{\answerYes}[1]{\textcolor{blue}{#1}} 
 
\newcommand{\answerNA}[1]{\textcolor{gray}{#1}}

\usepackage{newfloat}
\usepackage{listings}
\DeclareCaptionStyle{ruled}{labelfont=normalfont,labelsep=colon,strut=off} 
\floatstyle{ruled}
\newfloat{listing}{tb}{lst}{}
\floatname{listing}{Listing}
\title{Multi-agent Systems for Misinformation Lifecycle : Detection, Correction And Source Identification}
\author{
    Aditya Gautam
}
\affiliations{
    Independent Researcher \\
    Seattle, Washington, USA \\ 
    agautam7041@gmail.com
}

\begin{document}

\maketitle

\begin{abstract}
The rapid proliferation of misinformation in digital media demands solutions that go beyond isolated Large Language Model(LLM) or AI Agent based detection methods. This paper introduces a novel multi-agent framework that covers the complete misinformation lifecycle: classification, detection, correction, and source verification to deliver more transparent and reliable outcomes. In contrast to single-agent or monolithic architectures, our approach employs five specialized agents: an Indexer agent for dynamically maintaining trusted repositories, a Classifier agent for labeling misinformation types, an Extractor agent for evidence based retrieval and ranking, a Corrector agent for generating fact-based correction and a Verification agent for validating outputs and tracking source credibility. Each agent can be individually evaluated and optimized, ensuring scalability and adaptability as new types of misinformation and data sources emerge. By decomposing the misinformation lifecycle into specialized agents - our framework enhances scalability, modularity, and explainability. This paper proposes a high-level system overview, agent design with emphasis on transparency, evidence-based outputs, and source provenance to support robust misinformation detection and correction at scale.
\end{abstract}

%

\section{Introduction}

Recent research underscores the growing sophistication of LLMs in identifying and countering misinformation, while also revealing critical gaps in their reliability  \cite{wang2023assessingreliabilitylargelanguage} , bias \cite{lin-2025} and explainability \cite{cambria2024xaimeetsllmssurvey}. Studies by \citet{chen2023combatingmisinformationagellms} highlight the effectiveness of augmenting LLMs with external knowledge and tools for fact-checking, as well as fine-tuning and knowledge distillation. For instance, 
\citet{wang2024mmidrteachinglargelanguage} pioneered a knowledge distillation approach for multimodal misinformation detection, enhancing interpretability in complex image-text claims. Highlighting the benefits of multi-agent architectures, \citet{li2024largelanguagemodelagent} introduced FactAgent, breaking down fact-checking into specialized modules for evidence retrieval and source cross-referencing. To mitigate bias within these modules, \citet{borah2024implicitbiasdetectionmitigation} employed ensemble methods, leveraging self-reflection to reduce discriminatory task assignments. Building on collaborative detection, \citet{lakara2025madsherlock} proposed MAD-Sherlock, a debate-driven system that employs agent collaboration to reduce hallucinations and strengthen fact-verification, while \citet{tian2024webretrievalagentsevidencebased} integrated web-retrieval agents to boost detection over standalone models. Notably, \citet{minici2025iohuntergraphfoundationmodel} refined disinformation campaign detection with IOHunter, achieving high precision in orchestrated network identification. Finally, \citet{10817610} offered a comparative study on how LLMs handle political misinformation, revealing persistent challenges in grounding responses with credible sources. Collectively, these works confirm that LLM-driven techniques can significantly elevate detection capabilities for misleading content across multiple languages, modalities, and domains.In addition, \citet{tang2024minicheckefficientfactcheckingllms} demonstrated how subtle linguistic manipulation can degrade LLM-based fact-checking, showcasing lingering vulnerabilities in single-agent systems.

Although existing agentic systems often excel in specialized tasks, they typically concentrate on detection alone and fall short of covering the entire misinformation lifecycle, which is needed to fully understand the proliferation and the extend of misinformation. Here the lifecycle refers to understanding different type of misinformation in the given claim, correcting it with other authentic sources, verification with robust reasoning model, getting all other sources where the misinformation is present on same context, and understanding the root cause i.e initial misinformation source (which may or may not be the input claim). This is a unique value addition that the proposed framework provides. To address the above mentioned gaps, this work introduce a five specialized agents—data source authentication and indexing, multi-class classification, evidence ranking with confidence scores, correction generation with cross validation, and source-focused verification. By separating source provenance tracking from core analytical tasks and implementing cross-agent audit trails, our framework aims to achieves higher modularity and interpretability than tightly coupled, domain-specific models. This end-to-end architecture directly tackles transparency and adversarial resilience challenges, establishing a comprehensive solution for managing misinformation from start to finish. Through dynamic knowledge indexing, focused agent specialization, and explicit source audits, it aims to outperforms monolithic or single-agent systems in accuracy, adaptability, and clarity. By continuously tracing misinformation from its origin to final verification, the framework ensures timely and reliable oversight in an ever-changing information environment, culminating in a flexible yet complete multi-agent paradigm that holistically tracks misinformation, provides vetted evidence, and generates corrections with confidence scores. This system is envisioned for use by governmental bodies, fact-checking organizations, and media platforms to monitor and understand misinformation spread on specific topics (e.g., warfare, elections). By allowing users to define topics of interest (e.g., via keywords or BM25 filters), the system can identify and verify content against established ground-truth sources, tracing the lineage and origin of misinformation based on publication and modification timestamps in the metadata.

\section{Related work}

Recent advancements in multi-agent LLM frameworks demonstrate significant progress in addressing misinformation through collaborative architectures and overcoming the limitation with standalone LLMs. Hybrid systems, such as LLM-Consensus \citet{lakara2025llmconsensusmultiagentdebatevisual} , improve explainability and reduce hallucinations by combining multi-agent reasoning with external retrieval. MAD-Sherlock \citet{lakara2025madsherlock} introduces a debate-driven system where multi-modal agents assess contextual consistency in out-of-context (OOC) misinformation, achieving state-of-the-art accuracy without domain-specific fine-tuning by leveraging external retrieval and collaborative reasoning. Complementing this, MACAW framework by \citet{wu2024detecting} employs three specialized agents i.e. Retrieval, Detective and Analyst to cross-validate multi-granularity evidence, improving OOC detection accuracy through structured workflows. FactAgent by \citet{li2024largelanguagemodelagent} modularize fact-checking into evidence retrieval, temporal verification, and source cross-referencing, enhancing interpretability in veracity assessment. Expanding scope, \citet{tian2024webretrievalagentsevidencebased} integrate web retrieval agents with LLMs, boosting detection F1-scores compared to standalone models, while \citet{wang2023attackingfakenewsdetectors} expose vulnerabilities in graph-based detectors using adversarial multi-agent reinforcement learning. \citet{minici2025iohuntergraphfoundationmodel} advanced coordination detection with IOHunter, combining graph neural networks and LLMs to identify orchestrated disinformation campaigns. In this proposal, comprehensive and modular approach is taken to cater to all aspects of misinformation life-cycle i.e. origin, proliferation, detection, correction. This flexibility would allow this framework to be applied in any particular domain with plug and play.

\section{Proposed Multi-agents Framework}
The proposed architecture comprises five distinct agents working in a pipeline. First, the Classifier Agent analyzes the input claim and classifies it into a specific types of misinformation. This classification guides the Extractor Agent in querying the Indexer Agent's comprehensive database to retrieve relevant sources and their lineage. The Extractor Agent then ranks these sources based on authenticity, and alignment with the claim. The Corrector Agent, leveraging advanced reasoning capabilities, cross-validates the information and generates an accurate correction if misinformation is detected, along with supporting sources. Finally, the Verification Agent validates the outputs of the other agents to ensure the overall accuracy and coherence of the misinformation management process. The proposed multi-agent framework is designed for an end-to-end latency on the order of minutes. This accounts for the indexing pipeline for new content and the retrieval of relevant information for misinformation classification. While each agent operates within seconds to minutes, the overall system's timeliness is acknowledged as a critical factor, especially for sensitive topics like elections or conflicts. For inter-agent coordination, two design patterns are possible i.e. centralized and decentralized approaches. A centralized model would feature a master agent managing communication and policy enforcement, ensuring adherence to guidelines at the cost of potential latency. A decentralized model, where agents communicate directly, offers lower latency but requires robust mechanisms to prevent miscoordination. The choice of architecture will be guided by further research into the trade-offs between control, latency, and system resilience. A detailed discussion on the chosen orchestration design will be included in future work detailing the system's implementation.

\begin{itemize}
\item \textbf{Classifier Agent:} The Classifier Agent is designed to process any incoming content or article, which may or may not be misinformation. Its initial role is to perform a multi-class classification to detect if given context is misinformation or not, and if yes, of what types. This classification will be handled by a fine-tuned encoder-based model (e.g., RoBERTa) trained on a proprietary, internally labeled dataset curated by fact-checking teams. The Classifier Agent utilizes the text understanding and multi-class classification capabilities of LLMs. Its primary function is to analyze an input claim and categorize it into a predefined set of misinformation classes. This set includes categories such as statistical error, cherry-picking, propaganda, misrepresentation, historical manipulation, logical fallacy, factual error etc. Establishing a well-defined and comprehensive taxonomy of misinformation types is crucial for accurate classification and for guiding the subsequent actions of other agents, particularly the Extractor Agent. Classifier agent can have one or multiple LLM models with varying capabilities to detect misinformation classes, and use the ensembles based voting mechanism for boosting multi-classification performance. Carefully crafted prompt instructions along with fine-tuning on specific datasets can significantly influence the accuracy of the classification. The choice of LLMs itself is also critical, balancing accuracy requirements with computational cost. 

\item \textbf{Indexer Agent:} The Indexer Agent is responsible for indexing a wide range of data sources, ensuring that the indexed content is suitable for addressing the various types of misinformation identified by the Classifier Agent. This may involve utilizing LLM capabilities for text understanding to facilitate metadata extraction from the indexed content. The agent indexes web pages, news articles, scientific databases, statistical datasets, historical documents, fact-checking archives, etc based on the problem at hand. A diverse and comprehensive index is essential to provide the Extractor Agent with the necessary resources to verify different kinds of claims. New authentic data sources like Google Data commons etc. can be best leveraged by this agent along with official data from the government like data.gov and different organizations like WHO, UNESCO etc. The Indexer Agent ensures that the indexed content includes appropriate metadata, such as the source, publication date, topic, and a potential reliability score. This metadata enhances the searchability and relevance of the indexed data for the Extractor Agent. Possible indexing techniques include traditional keyword-based indexing, and more advanced semantic indexing using embeddings generated by language models like Retrieval Augmented Generation (RAG). Establishing ground truth for misinformation detection involves several approaches. This agent is also responsible for generating metadata like authenticity score, topics, category etc based on the data-source description, usage, columns structures and other information like data source site description, usage, research citations etc. Additionally, developing automated methods for ground truth generation and validation, potentially by cross-referencing information across multiple highly reliable sources, is crucial for scaling up misinformation detection efforts.The Indexer Agent employs a multi-step pipeline that includes data cleaning, standardization of various content formats (HTML, PDF etc.) into a uniform structure with rich metadata (title, author, domain, links etc.). Content is then segmented using appropriate chunking strategies (e.g., semantic, fixed-size) optimal for the data and task. These chunks are converted into vector embeddings using fine-tuned models and stored in a vector database (e.g., using FAISS or similar technologies) to enable efficient (O(1) retrieval time) top-K similarity searches. This scalable architecture is standard in industry for RAG systems.

\item \textbf{Extractor Agent:} The Extractor Agent leverages LLM capabilities in text understanding, information extraction, semantic similarity assessment, and ranking 9. It receives the misinformation type classification from the Classifier Agent and uses this information to refine its search strategy within the indexed data. Based on the classification, the Extractor Agent prioritizes querying specific sections of the indexed data or dedicated databases that are most relevant to the identified misinformation type. For instance, if the claim is classified as a statistical error, the agent will focus its queries on statistical databases and reports. For example, If the claim is a recent news article about some politicians or celebrity, it would fetch all the news articles related to this, and the statement made by the actors in context. 
Furthermore, the Extractor Agent adjusts its re-ranking strategy for extracted documents to prioritize those most pertinent to the specific type of misinformation. For historical misrepresentation, primary historical sources and expert analyses will be given higher priority. The agent retains its original functionalities of assigning authenticity scores to sources, potentially based on traditional information retrieval and ranking features like domain reputation, page rank, word frequency analysis, author reputation etc. It also continues to determine the alignment of the extracted sources with the input claim using semantic similarity measures and to identify the lineage or original source of the input information based on publication timestamps and content similarity.

\item \textbf{Corrector Agent:} The Corrector Agent utilizes a strong reasoning LLM to take inputs from the Extractor Agent (claim, sources with authenticity scores, lineage etc.) and misinformation categories as identified by the Classifier Agent along with confidence score. Based on these inputs, it performs in-depth research and cross-validation using the extracted sources, and conditionally conducting additional searches i.e. encompasses both querying the internal, pre-indexed database more extensively (e.g., by retrieving a larger number of context documents) and performing external online searches via integrated tools (e.g., Google Search APIs) or querying other accessible external databases to gather further validating evidence. The agent then generates accurate corrections tailored to the specific type of misinformation identified. For example, if the misinformation is a statistical error, the Corrector Agent will aim to provide the correct statistical information along with its source, re-think about potential other sources to cross validate it, and append different citations based on the authenticity score or other pre-defined criteria. The Corrector Agent also provides reliable information for all the sources with metadata like timestamps etc. to ensure a lineage of misinformation and source identification can be done.  

\item \textbf{Verification Agent:}
The Verification Agent is responsible for validating the outputs of the other agents against predefined criteria 3. These criteria may include logical consistency, adherence to source reliability thresholds, format specification, tone and the alignment between the generated correction and the identified misinformation type. The primary goal of this agent is to ensure the overall coherence and accuracy of the misinformation detection and correction process. It acts as a final quality check, mitigating potential errors or biases that may have been introduced by the other agents. The Verification Agent may utilize LLM capabilities for reasoning and text understanding to perform these validation checks. This is where human-in-loop would be best utilized to understand the performance of the human based verification and labeling and through this multi-agent framework. Verification agents can be potentially merged with Corrector agents to do verification and correction, depending on the use case,domain and complexity of the problem. This agent cross validate any additional information or subjective instructions provided by the users like tone, number of doctors for cross checking, presentation format etc. and prepare the final response and potentially use different tools at the disposable to generate reports, add lineage information to spreadsheet, display it through charts and diagrams etc.
\end{itemize}

\subsection{Advantages of the Proposed Framework}

\begin{itemize}
\item \textbf{Systematic Evaluation:} Each agent can be independently monitored and fine-tuned, allowing focused performance assessments across different tasks.
\item \textbf{Freshness:} A dynamically updated index ensures real-time adaptation to newly emerging misinformation patterns and sources.
\item \textbf{Adaptability:} Modular design lets practitioners add or revise agents (e.g., introducing new data sources) without overhauling the entire system.
\item \textbf{Specialization:} Each agent targets a distinct task—indexing, classification, extraction, correction, or verification— instead of “jack-of-all-trades” model.
\item \textbf{Cost Optimization:} Resource usage is allocated based on agent complexity, minimizing overall computational overhead.
\item \textbf{Knowledge Sharing:} Agents share findings through a unified communication layer, allowing seamless collaboration and evidence cross-referencing.
\item \textbf{Strength Maximization:} Researchers can focus on upgrading individual components, such as classification or correction, without destabilizing the entire pipeline.
\item \textbf{Robustness and Reliability:} The decomposition of tasks into specialized agents reduces single points of failure and improves error detection and correction at each stage.
\end{itemize}

\begin{table*}[ht]
\centering
\small
\begin{tabular}{|p{0.08\textwidth}|p{0.22\textwidth}|p{0.22\textwidth}|p{0.22\textwidth}|}
\hline
\textbf{Agent} & \textbf{Functionality} & \textbf{Key Benefits} & \textbf{LLM Capabilities} \\
\hline
Classifier & Initial claim categorization i.e. misinformation categories, topic, sentiment & Provides high-confidence triggers to extract sources and grounds for other agents & Text understanding, multi-class classification, few-shot/zero-shot learning \\
\hline
Indexer & Database indexing, finding new sources, embedding and chunking & Efficiently categorizes information with source authenticity score & Text understanding and metadata generation based on data sources \\
\hline
Extractor & Information retrieval, source detection (timestamp \& metadata analysis, lineage tracing, confidence score) & Offers comprehensive understanding of misinformation origin and spread & Text understanding, information extraction, semantic similarity, ranking, targeted querying \\
\hline
Corrector & Misinformation correction generation with cross validation from additional sources & Produces accurate and contextually relevant corrections & Strong reasoning, knowledge generation, cross-validation, in-context learning \\
\hline
Verification & Final veracity determination, formatting, comprehensive reports generation etc. & Ensures coherence, accuracy, and consistency of the overall assessment & Reasoning, text understanding, fact verification, logical consistency checks \\
\hline
\end{tabular}
\caption{Characteristics and Functional Capabilities of Different Agents}
\end{table*}

\section{Discussion}

The proposed LLM-based multi-agent framework offers a structured and scalable approach to managing the full misinformation lifecycle—from classification and detection to correction and source verification. Unlike monolithic models that attempt to handle all tasks within a single architecture, this system distributes responsibilities across specialized agents, improving transparency, explainability, and robustness. The modular design allows for independent upgrading of agents (e.g., improving the Extractor Agent with more advanced retrieval models) without overhauling the entire system. This enhances adaptability to emerging misinformation patterns and evolving data sources. Moreover, by integrating real-time evidence retrieval, citation generation, and reasoning capabilities, the framework not only identifies misinformation but also corrects it with traceable justification—critical for user trust and accountability. However, challenges remain, including managing inter-agent coordination, mitigating latency introduced by multi-stage processing, and ensuring reliability of indexed data. Future iterations could benefit from incorporating self-evaluation loops and reinforcement learning to dynamically improve agent collaboration and performance. There are several other things that needs to very well thought through in Multi-agent system like agents coordination protocols, collusion, policy violation, bias amplification etc,  Even though the system paves the way for more resilient and interpretable management of misinformation, there are some risk as mentioned in detail by \citet{hammond2025multiagentrisksadvancedai} and complexity that comes with multi-agent systems as mentioned in \citet{cemri2025multiagentllmsystemsfail}. Some of the challenges in the proposed framework:

\begin{itemize}
\item \textbf{Data Quality and Freshness:} The effectiveness of the Indexer Agent is contingent upon the quality and recency of the underlying data sources.
\item \textbf{Coordination and Collusion:} Ensuring seamless coordination among agents is complex, especially as the number of agents increases. Poor coordination can lead to conflicts or redundant actions. Additionally, there's a risk of agents colluding, intentionally or unintentionally.
\item \textbf{Cost and Scalability:} The cost considerations for continuous indexing, complex reasoning, and ensemble verification could be substantial, particularly for large-scale deployment.
\item \textbf{Network Effects:} The interdependent nature of agents means that the behavior of one agent can influence others, sometimes leading to unintended consequences or emergent behaviors that are difficult to predict and control.
\item \textbf{Security and Privacy:} MAS often involve extensive data sharing among agents, raising concerns about data privacy and security. Unauthorized access or data breaches can lead to significant vulnerabilities, especially when agents operate across different platforms or organizations
\end{itemize}

\section {Future work}
This paper only proposed the multi-agent framework for misinformation lifecycle in this paper as the foundational system architecture. Future research will focus on implementing and empirically evaluating the proposed multi-agent framework using benchmark misinformation datasets such as FakeNewsNet \cite{shu2018fakenewsnet}, and WELFake \cite{9395133}. We acknowledge the complexity of deploying such a system, which necessitates robust infrastructure including real-time indexing, scalable databases, and comprehensive data scraping capabilities to build and maintain a reliable ground-truth repository. The current paper presents a high-level framework, and detailed implementation specifics, including the orchestration of these components, are part of ongoing and future development. Experiments will assess each agent’s performance individually and in pipeline mode, measuring metrics like accuracy, precision, recall, latency, and explainability. Additionally, ablation studies can explore the impact of agent specialization and external retrieval on system robustness. Future extensions may also incorporate multilingual capabilities, cross-modal misinformation detection, and user feedback loops for adaptive learning and continuous improvement.

\bibliography{aaai25}

\subsection{Ethical Consideration}
While this paper primarily proposes a conceptual multi-agent framework for addressing the misinformation lifecycle and outlines high-level implementation aspects without presenting experimental data, it is crucial to acknowledge the inherent ethical considerations. The future development and deployment of such a system demand careful scrutiny of each specialized agent. For instance, the Indexer Agent must address potential biases in the selection and maintenance of "trusted" repositories. The Classifier Agent's categorization of misinformation types requires safeguards against mislabeling and perpetuating harmful stereotypes. The Extractor Agent's evidence retrieval and ranking mechanisms must ensure fairness and avoid amplifying certain viewpoints disproportionately. Furthermore, the Corrector Agent's generation of corrections carries the responsibility of accuracy and neutrality, while the Verification Agent's processes for validating outputs and source credibility must be transparent and robust against manipulation. Future work will need to rigorously evaluate and mitigate these and other ethical challenges to ensure responsible application of this framework. Below are the ethical checklist responses for this paper.

\begin{enumerate}

\item For most authors...
\begin{enumerate}
    \item  Would answering this research question advance science without violating social contracts, such as violating privacy norms, perpetuating unfair profiling, exacerbating the socio-economic divide, or implying disrespect to societies or cultures?
    \answerYes{Yes}
  \item Do your main claims in the abstract and introduction accurately reflect the paper's contributions and scope?
    \answerYes{Yes}
   \item Do you clarify how the proposed methodological approach is appropriate for the claims made? 
    \answerYes{Yes}
   \item Do you clarify what are possible artifacts in the data used, given population-specific distributions?
    \answerNA{NA, No data is used in this paper}
  \item Did you describe the limitations of your work?
    \answerYes{Yes, there is a section on potential limitation}
  \item Did you discuss any potential negative societal impacts of your work?
    \answerYes{Yes, though this is little beyond the scope, it is mentioned that the agents and datasources needs to be well vetted and the labeling would be done by experts and authorized body to avoid any societal bias and adverse impact}
      \item Did you discuss any potential misuse of your work?
    \answerNA{No, this is the beyond the scope of the framework proposed}
    \item Did you describe steps taken to prevent or mitigate potential negative outcomes of the research, such as data and model documentation, data anonymization, responsible release, access control, and the reproducibility of findings?
    \answerNA{NA, No data is used}
  \item Have you read the ethics review guidelines and ensured that your paper conforms to them?
    \answerYes{Yes, I have read them}
\end{enumerate}

\item Additionally, if your study involves hypotheses testing...
\begin{enumerate}
  \item Did you clearly state the assumptions underlying all theoretical results?
    \answerNA{NA, there is no experimentation in this paper}
  \item Have you provided justifications for all theoretical results?
    \answerNA{NA}
  \item Did you discuss competing hypotheses or theories that might challenge or complement your theoretical results?
    \answerNA{NA}
  \item Have you considered alternative mechanisms or explanations that might account for the same outcomes observed in your study?
    \answerNA{NA}
  \item Did you address potential biases or limitations in your theoretical framework?
    \answerYes{Yes, it has been discussed in one section}
  \item Have you related your theoretical results to the existing literature in social science?
    \answerNA{NA, there is no relevance of social science}
  \item Did you discuss the implications of your theoretical results for policy, practice, or further research in the social science domain?
    \answerNA{NA}
\end{enumerate}

\item Additionally, if you are including theoretical proofs...
\begin{enumerate}
  \item Did you state the full set of assumptions of all theoretical results?
    \answerNA{NA}
	\item Did you include complete proofs of all theoretical results?
    \answerNA{NA}
\end{enumerate}

\item Additionally, if you ran machine learning experiments...
\begin{enumerate}
  \item Did you include the code, data, and instructions needed to reproduce the main experimental results (either in the supplemental material or as a URL)?
    \answerNA{NA, no experimentation in this proposal just the framework}
  \item Did you specify all the training details (e.g., data splits, hyperparameters, how they were chosen)?
    \answerNA{NA}
     \item Did you report error bars (e.g., with respect to the random seed after running experiments multiple times)?
    \answerNA{NA}
	\item Did you include the total amount of compute and the type of resources used (e.g., type of GPUs, internal cluster, or cloud provider)?
    \answerNA{NA}
     \item Do you justify how the proposed evaluation is sufficient and appropriate to the claims made? 
    \answerNA{NA}
     \item Do you discuss what is ``the cost`` of misclassification and fault (in)tolerance?
    \answerNA{NA}
  
\end{enumerate}

\item Additionally, if you are using existing assets (e.g., code, data, models) or curating/releasing new assets, \textbf{without compromising anonymity}...
\begin{enumerate}
  \item If your work uses existing assets, did you cite the creators?
    \answerNA{NA}
  \item Did you mention the license of the assets?
    \answerNA{NA}
  \item Did you include any new assets in the supplemental material or as a URL?
    \answerNA{NA}
  \item Did you discuss whether and how consent was obtained from people whose data you're using/curating?
    \answerNA{NA}
  \item Did you discuss whether the data you are using/curating contains personally identifiable information or offensive content?
    \answerNA{NA}
\item If you are curating or releasing new datasets, did you discuss how you intend to make your datasets FAIR (see \citet{fair})?
\answerNA{NA}
\item If you are curating or releasing new datasets, did you create a Datasheet for the Dataset (see \citet{gebru2021datasheets})? 
\answerNA{NA}
\end{enumerate}

\item Additionally, if you used crowdsourcing or conducted research with human subjects, \textbf{without compromising anonymity}...
\begin{enumerate}
  \item Did you include the full text of instructions given to participants and screenshots?
    \answerNA{NA}
  \item Did you describe any potential participant risks, with mentions of Institutional Review Board (IRB) approvals?
   \answerNA{NA}
  \item Did you include the estimated hourly wage paid to participants and the total amount spent on participant compensation?
    \answerNA{NA}
   \item Did you discuss how data is stored, shared, and deidentified?
   \answerNA{NA}
\end{enumerate}

\end{enumerate}

\end{document}